\documentclass{amsart}
\usepackage{amsmath,graphicx,url,times}
\usepackage{amssymb}
\usepackage{times}
\usepackage{tikz}
\usetikzlibrary{arrows,automata}

\usepackage{textcomp}
\usepackage{amsmath, amsfonts, amsthm, amssymb, amscd}
\usepackage{caption, subfigure}
\usepackage{fixltx2e}

\usepackage{ulem,soul}	

\title[FD2L: sparse reconstruction of the topology]{From dynamics to links: a sparse reconstruction of the topology of a  neural network}

\begin{document}

\author[G. Aletti]{Giacomo Aletti}
\address{ADAMSS Center, Universit\`a degli Studi di Milano, 20131 Milano, Italy}
\email{giacomo.aletti@unimi.it}

\author[D. Lonardoni]{Davide  Lonardoni}
\address{Neuroscience Brain Technology, Istituto Italiano di Tecnologia, via Morego 30, 16163 Genova, Italy}

\author[G. Naldi]{Giovanni Naldi}
\address{ADAMSS Center, Universit\`a degli Studi di Milano, 20131 Milano, Italy}

\author[T. Nieus]{Thierry Nieus}
\address{Neuroscience Brain Technology, Istituto Italiano di Tecnologia, via Morego 30, 16163 Genova, Italy}



\begin{abstract}
One major challenge in neuroscience is the identification of interrelations between signals reflecting neural activity and  how information processing occurs in the neural circuits. At  the cellular and molecular  level, mechanisms of signal transduction   have been studied intensively and a better knowledge and understanding of some basic processes of information handling by neurons has been achieved.   In contrast, little is known about the organization and function of complex neuronal networks. Experimental methods are now available to simultaneously monitor electrical activity of a large number of neurons in real time.  Then, the qualitative and quantitative analysis of the spiking activity of individual neurons is a very valuable tool for the study of the dynamics and architecture of the neural networks. Such activity is not due to the sole intrinsic properties of the individual neural cells but it is mostly consequence of the direct influence of other neurons.  
The deduction of the effective connectivity between neurons, whose experimental spike trains are observed, is of crucial importance in neuroscience: first for the correct interpretation of the electro-physiological activity of the involved neurons and neural networks, and, for correctly relating the electrophysiological activity to the functional tasks accomplished by the network.
In this work we propose a novel  method for the identification of connectivity  of neural networks using recorded voltages. Our approach is based on the assumption that the network has a topology with sparse connections. 
After a brief description of our method we will report the performances and compare it to the cross-correlation computed on the spike trains, that represents a gold standard method in the field. 
\end{abstract}

\maketitle

\section{Introduction}
\label{sec:intro}
Along the latests years there have been enormous progresses in the Neuroscience field
that have revolutionized the way we envisage the brain functions.
In this regard, the technological advancements have been fundamental to improve the recording capability
from brain areas and neural populations \cite{Stevenson2011}.
Nowadays, multi-site recordings can be achieved from thousands of channels (sites) with a good spatial (at the cellular level) and temporal resolution (less than one millisecond for the action potential) yielding a good description of the underlying network dynamics. Given that the brain operates on a single trial basis such recordings are instrumental to understand the neural code \cite{Churchland2007}.
As a first step, multi-site recordings allow to quantify the information flow in the network. The anatomical wiring (i.e. Structural Connectivity, SC) clearly plays a fundamental role to understand how cells comunicate among them but it is often not well known neither it can by itself explain the overall network activity.
Multi-site recordings can be used to infer statistical dependencies (i.e. Functional Connections, FC) among the recorded units and to track the information flow in the network \cite{Bullmore2009}.
On the other hand the Effective Connectivity (EC) denotes the directed causal relationship between the recorded sites. The EC is typically estimated by stimulating one cell and studying the effects on the connected elements. Alternatively the EC can also be studied establishing a causal mathematical model between the recorded units data.

Importantly, multi-site recordings raise some limitations that need to be evaluated carefully before any further analysis. First, the experimental sessions are often severly limited in time. Second, the high dimensional data sets involve a set of numerical and mathematical problems that would be hard to face even with long enough recording sessions. These issues are common to different fields and have been coined as 'curse of dimensionality'. To overcome these issues, two approaches are typically foreseen. A first solution consists into a reductionist approach that projects the data on a lower dimenstional space that can better elucidate the underlying processing.
Another possibility consists into fitting the observed data to a low-dimensional model that captures the salient properties of the dynamics \cite{Stevenson2011}.
Here we introduce a model of effective connectivity that gets rid of the dimensionality problem by introducing a natural constrain of almost all biological networks: the sparseness among the connected units (see e.g.\ \cite{Causal11}).
Other approaches, such as the multivariate autoregressive model \cite{Winterhalder2005}, allow for the possibility of a fully connected network in which every node may influence all the other nodes. 
However this is somewhat unrealistic for biological networks (i.e. each neuron is directly influenced by only a small subset of neurons) and it also leads to practical challenges.
In fact, most of the connectivity tools used to understand the communication among neuronal populations are based on linear models although it is widely recognized that the interactions (i.e. synaptic currents) among neural cells are nonlinear. Then, fully connected networks involve models that can easily be over-parameterized and hard to solve.
Therefore current methods are not well suited to robustly infer on the network connectivity from the time series in different contexts.   
Among these approaches, the Granger  causality \cite{Granger1969,Bressler2011} is probably the most prominent and most widely used concept. 
This concept of causality does not rely on the specification of a scientific model and thus is particularly suited for empirical investigations of cause-effect relationships. 
On the other hand, if important relevant variables are not included in the analysis, the Granger causality can lead to so-called spurious causalities . 
In order to capture nonlinear interactions between even short and noisy time series, we consider
an event-based  model. Then, we involve the physiological basis of the signal, which is likely non-linear. 

In Section~\ref{sec:probForm} we introduce a general setting for the problem. In Section~\ref{sec:expSett} we describe our experimental
settings and data preprocessing. The results are reported in Section~\ref{sec:expRes}, while final remarks are postponed in the last section.

\section{Problem Formulation}\label{sec:probForm}

\subsection{Network representation of the problem}
\label{ssec:PF1}
\begin{figure}[htb]
\centering
\begin{tabular}{cl}
\begin{tikzpicture}[->,>=stealth',shorten >=1pt,auto,node distance=1.8cm,
                    semithick, baseline]
  \tikzstyle{every state}=[fill=blue,draw=none,text=white]

  \node[state]         (A)                    {$1$};
  \node[state]         (B) [above right of=A] {$2$};
  \node[state]         (D) [below right of=A] {$3$};
  \node[state]         (C) [below right of=B] {$4$};
  \node[state]         (E) [right of=C]       {$5$};

  \path (A) edge              node {} (B)
            edge              node {} (C)
           (B) edge              node {} (C)
        (C) edge              node {} (D)
            (D) edge              node {} (A)
		edge  [bend right]  node {} (E)
        (E) edge [bend left]  node {} (A)
            edge   node {} (C);
\end{tikzpicture}
&
$
C_E =
\begin{pmatrix}
0 & 1 & 0 & 1 & 0 \\
0& 0 & 0 & 1 & 0 \\
1 & 0 & 0 & 0 & 1 \\
0& 0 & 1 & 0 & 0 \\
1 & 0 & 0 & 1 & 0
\end{pmatrix}
$
\end{tabular}
\caption{Example of a nework of $5$ neurons and its adjacency matrix.} \label{fig:network}
\end{figure}
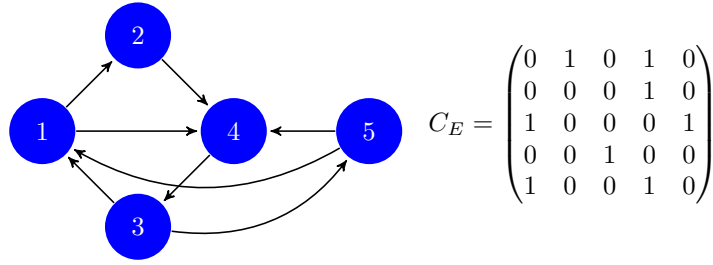

Let us consider a graph $G = (V,E)$, where $V$ corresponds to a set of $N$ neurons connected with a directed graph given by the edges contained in $E $ of ordered pairs of vertices contained in $ V \times V$. 
On each node $i \in V$, two stochastic processes are observed:
\begin{itemize}
\item an external process $\{X_i(t), 0\leq t \leq T\}$ (spike);
\item an internal process $\{Y_i(t), 0\leq t \leq T\}$.
\end{itemize}
We assume a \emph{Local Markov Property (LMP)}: the relevant information on each process $Y_i(t),i=1,\ldots, N$ is ``contained'' 
on the external  activity of the neurons $j$ connected to it, and in a time interval $\delta$. For example, in the network given in Figure~\ref{fig:network}, the law of the $4$th neuron depends on the external activity of the $1$st, $2$nd, and $5$th neuron together with its own internal activity.

We explain now the LMP in terms of the $\sigma$-algebras of events observable by the processes.
We denote by $\mathcal{F}_{t^-} $ the observable events of the ``past'' of the external processes until time $t$, and it will be defined as the $\sigma$-algebra of events generated by the processes $\{X_i(s), s<t, i \in V\}$:
\(
\mathcal{F}_{t^-}  = \bigvee_{i \in V} \sigma(X_i(s), s< t ) .
\)
The $\sigma$-algebra $\mathcal F^{(i)}_{t^-} $ contains the observable events of the ``close past'' of the in-neighborhoods of the $i$-th neuron:
\(
\mathcal F^{(i)}_{\delta,t^-} = \bigvee_{j\colon (j,i)\in E} \sigma(X_j(s), t-\delta \leq s< t ) .
\)
Moreover, we will denote by $\mathcal{G}_{t^+}$ the observable events of the ``future'' of the internal processes:
\(
\mathcal{G}^{(i)}_{t^+}  = \sigma(Y_i(u), t \leq u ).
\)

With this notation, LMP says that the future of the $i$-th internal process and past of all the external processes are conditionally independent given the close past of the $i$-th neuron:
\begin{equation}\label{eq:LMP}\tag{LMP}
P( G | \mathcal{F}_{t^-} ) =
P( G | \mathcal{F}^{(i)}_{\delta,t^-} ) , \qquad \forall G\in \mathcal{G}^{(i)}_{t^+} , \forall i.
\end{equation}
In other words, $\mathcal{F}^{(i)}_{\delta, t^-} $ gives all the relevant information on the future of the internal activity of the $i$-th neuron.

\subsection{Event-based discrete time model}

We use a particular model of the class of processes satisfying \eqref{eq:LMP}.
We sample the processes at discrete times $t_m = m \delta$, and we
assume an \emph{event-based model}: both the internal and external processes are simple point processes,
and we assume that at most one event may occur for each process in any time interval.
We note that the definition of event depends on the knowledge of the investigator and on the available data.
For the purpose of this paper, the definition of events will be introduced in the Section~\ref{sec:preproc}.
Given an event-based model, for any $m=1, \ldots, M$, we observe
\[
y_i^m =
\begin{cases}
+1 & \text{if there has been an event for } Y_i(s) \text{ duting }(t_{m-1},t_{m}];\\
0 & \text{otherwise};
\end{cases}
\]
and
\[
x_i^m =
\begin{cases}
+1 & \text{if there has been an event for } X_i(s) \text{ duting }(t_{m-1},t_{m}];\\
0 & \text{otherwise};
\end{cases}
\]
The model can be completely characterized  by defining the family of conditional probabilities:
for any $i,m$,
\[
\pi (i,m) = P( y_i^{m+1} = 1 | x_j^l, l\leq m, j\in V) = P( y_i^{m+1} = 1 | x_j^m, (j,i)\in E)  
\]
the last equality being a consequence of \eqref{eq:LMP}.

To model the association among the nodes, we assume in this paper
a time-homogeneous linear logit model:
\begin{equation}\label{eq:THLLm}
\pi(i,m) = ( 1 + \exp( - \beta_0 - \sum_{ (j,i)\in E} \beta_{j,i} x_j^m ) )^{-1} 
\end{equation}
Given a set $\{\omega_{i,m}\}$ of positive real numbers, the negative weighted log-likelihood function for our processes reads:
\begin{equation}\label{eq:nLL}
\ell_M(\boldsymbol{\beta}) = - \sum_{m=1}^{M-1}\sum_{i\in V} \omega_{i,m}
\Big( y_i^{m+1} \log \pi(i,m) + (1-y_i^{m+1}) \log (1-\pi(i,m))\Big)
\end{equation}
Note that the significant parameters  $\{\beta_{j,i}, (j,i)\in E\}$ of our model depend on the topology of the network.
The choice of the weights is done here to balance the number of $y_i^{m} = 0$ and $y_i^{m} = 1$. A simple choice might be
\( \omega_{i,m} = \sum_{j,l} y_j^{l} \) if $y_i^m = 0$ and \( \omega_{i,m} = \sum_{j,l} (1 -y_j^{l}) \) if $y_i^m = 1$.

\section{Experimentimental settings}\label{sec:expSett}

\subsection{Simulations \label{simul}}
We simulate the potential of the $j$-th neuron $V_j(t)$ by the Hodgkin\&Huxley equations of the cerebellar granule cell (GrC, \cite{DAngelo2001}). Heterogeinity among the GrCs is introduced by randomizing the resting potentials with a current of $2\pm0.2$ pA.
The network consists of 20 neurons with a connectivity probability of 0.3 and all connections are directional (i.e. chemical synapses). Since self-connections (i.e. 'autapses') are not so frequent in the brain, they were not included in our network. Based on these numbers, the simulated network comprises on average 114 connections (i.e. $20\cdot19\cdot0.3$).
The chemical synapses are modeled by the fast AMPA excitatory currents \cite{Nieus2006}. In addition, to mimic a biologically realistic noisy regime, Poisson distributed AMPA currents (of frequency 0.2 Hz) are injected to all GrCs. 
The cellular and synaptic models are described in the literature \cite{DAngelo2001,Nieus2006} and the parameter settings and their changes respect to the literature are reported in the Table~\ref{tab:sett}.
\begin{table}[htb]
\begin{tabular}{lllll}
 \hline
 synaptic input &  parameter	   &  unit	 & value  &  value in literature \cite{Nieus2006} \\
 	AMPA			& gmax & pS & 800 & 1200 \\
 				& $r_1$	& $ms^{-1} mM^{-1}$   & 5.4 &   \\
 				& $r_2$ & $ms^{-1}$ & 0.84 &  	 \\	
 				& $r_6$	& $ms^{-1} mM^{-1}$ & 0 & 1.12 	  \\	
 	NMDA 		& not included here & & & \\    	
 				& & & &  \\
 synaptic noise 	& gmax & pS & 500  & 					\\
 				& $r_1$	& $ms^{-1} mM^{-1}$  & 5.4 &		\\
 				& $r_2$ & $ms^{-1}$  & 0.1 &		 \\	
 				& $r_6$	& $ms^{-1} mM^{-1}$   & 0  &	  \\	
 \hline
\end{tabular}
\caption{Values for the parameter settings for the dynamics of AMPA, NMDA and synaptic currents}\label{tab:sett}
\end{table}
Since here we are interested in highlighting the potential impact of our approach we discarded the contribution of other synaptic conductances that will be included in a next work.

\begin{figure}[htb]
\begin{center}
\includegraphics[width=.65\linewidth]{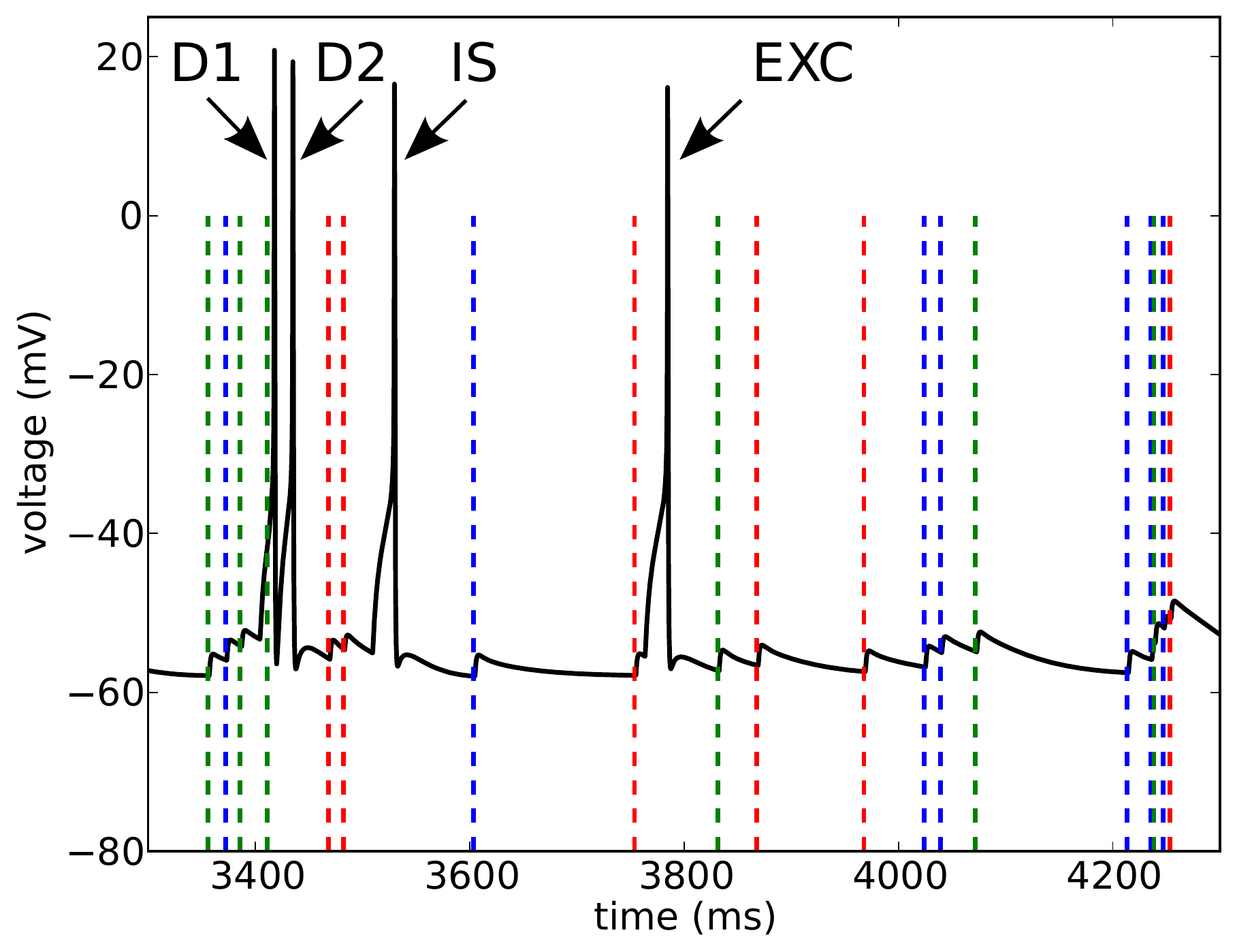}
\caption{Voltage trace of a neuron in the network. The voltage (black trace) is overlayed with the time stamps of the inputs (colored dash lines) to the cell. The voltage changes are clearly correlated with the input. However spikes D1, D2, IS, EXC are not apparently determined solely by the input. Note that the  blue, red and green colors correspond to three different cells.
}
\label{fig:FigRumore}
\end{center}
\end{figure}

We simulate five seconds of activity of such a network. A snapshot of the activity of a neuron in the network is shown in Figure \ref{fig:FigRumore}. The upswings of the potential are mainly determined by the inputs from the other cells (the time occurrence of these inputs is highlighted by the red, blue and green dashed lines). Interestingly, the spiking activity is not strictly determined by the input itself. In fact, the spike doublet (D1,D2), the isolated spikes (IS) and the spike from excitation (EXC) are not apparently determined by the input. 
Spike D1 arises from a depolarization that is not directly caused by the input and spike D2 can either by determined by high membrane excitation as well as by noise. The spike IS is most probably caused only by noise and spike EXC is again due to membrane excitation or simply by synaptic noise.

\subsection{Data preprocessing}\label{sec:preproc}

\begin{figure}[hbt]
\begin{center}
\includegraphics[width=.45\linewidth]{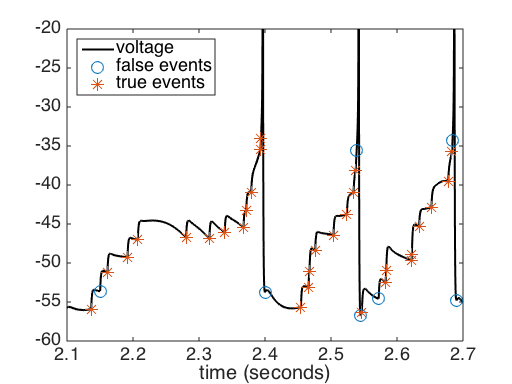}
\includegraphics[width=.45\linewidth]{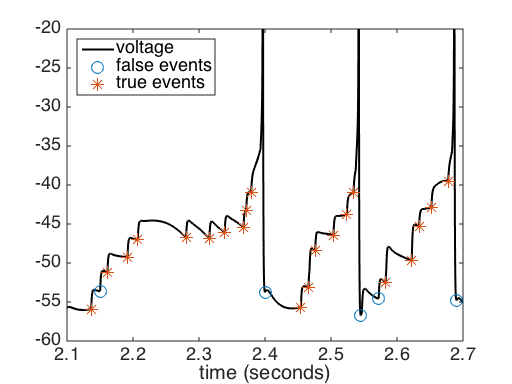} \\
\includegraphics[width=.45\linewidth]{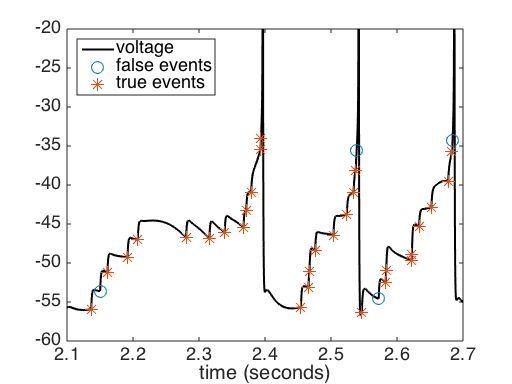}
\includegraphics[width=.45\linewidth]{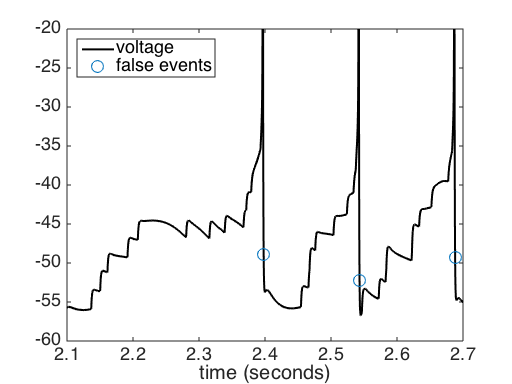}
\caption{Example of identification of events based on the voltage activity.
Events that are  {determined} by the links of the network are marked with $*$ {(i.e. true positive)}, while $\circ$ shows false identification
{(i.e. false positive)} (due, for example, to external noise).
Top left: identification of events with all the three factors of data preprocessing. Top right: identification of events with only (i) factor.
Bottom left: identification of events with only (ii) factor. Bottom right: identification of events with only (iii) factor.}\label{fig:preproc}
\end{center}
\end{figure}

The process $\{x_j^m, m=1,\ldots,M\}$  is the discretization (in time) of the spike activity of the $j$-th neuron, so that
$x_j^m=1$ if there has been a spike during the time interval $(t_{m-1},t_{m}]$.

The process $\{y_j^m, m=1,\ldots,M\}$ is a reaction process, nonlinear filter of the potential activity $V_j(t)$.
Here, $y_j^m=1$ if there has been an event for $V_j(t)$ during the time interval $(t_{m-1},t_{m}]$,
where an event at time $t$ depends on three factors (see Figure~\ref{fig:preproc}):

{\renewcommand*{\theenumi}{\roman{enumi}}
\begin{enumerate}
\item the right derivative $V_j'(t^+)$ must be greater than a positive threshold (increasing of potential after excitation);
\item the increasing of derivatives $V_j'(t^+) - V_j'(t^-)$ must be greater than a positive threshold (convex effect of potential due to excitation);
\item the left derivative $V_j'(t^-)$ must be greater than a negative threshold 
(in connection with other conditions, this avoid an event caused by the recover of the resting potential during the hyperpolarization phase; 
alone, this identifies those events).
\end{enumerate}}
More precisely, when a sequence of times of length $n\geq 1$ satisfies the requested conditions, an event is detected. The first time of
this sequence is the time of the event.

\subsection{Topology estimator} Given the data of our processes, we are interested in reconstructing the topology of our network. We want to give
an estimator $\hat{V}$ of the set $V$ of the edges.
In our contest, the edge $(j,i)$ exists when the process $y^m_i$ is directly caused by $x^m_j$, in the sense of \eqref{eq:LMP}.
In other words, when $\beta_{j,i}$ is different from zero in \eqref{eq:THLLm}. Therefore, we adopt the following strategy:
\begin{itemize}
\item
with a penalization technique, we find a sparse estimator $\hat{\boldsymbol{\beta}}$ of $\boldsymbol{\beta}$;
\item
we say that $(j,i)\in \hat{V}$ if $\hat{\beta}_{j,i}$  is different from zero.
\end{itemize}
In this paper, we adopt a $\ell_1$-penalization on the regression coefficients $\{{\beta}_{j,i},i,j\in E\}$. Given a 
positive penalization parameter $\lambda$,
let
\[
L(\boldsymbol{\beta}, \lambda) = \ell_M(\boldsymbol{\beta}) + \lambda \sum_{i,j\in E} |{\beta}_{j,i}|,
\]
and define the Lasso estimator
\(
\hat{\boldsymbol{\beta}}(\lambda) = \arg\min_{\boldsymbol{\beta}} L(\boldsymbol{\beta}, \lambda).
\)
As stated above, we define $\hat{V}(\lambda)$ as
\begin{equation}\label{eq:Vhat}
(j,i)\in \hat{V} \iff \hat{\beta}_{j,i}(\lambda)>0.
\end{equation}

Here we also compare the performances of the novel methodology with the standard cross-correlation that is widely used in the multi electrode array field \cite{Garofalo2009,Maccione2012,Ullo2014}.
The cross-correlation functions among the discrete spike trains are defined as:
\begin{equation}\label{eq:CrossCorr}
CC_{i,j}(\tau)=\frac{\langle ST_i \cdot ST_j(\tau)\rangle }{\sqrt{N_i\cdot N_j}}
\end{equation}
where $ST_i$, $ST_j$ are the binned spike trains (bin size 1 ms) and $N_i$,$N_j$ the corresponding number of spikes. The strength of a connection, between the nodes $i$ and $j$, is then given by the peak of the quantity $CC_{i,j}$ 
given in \eqref{eq:CrossCorr}. 
The network topology can then be inferred by retaining the strongest and most significants cross-correlation peaks that are overcome a selected threshold.  

\section{Experimental Results}\label{sec:expRes}
\begin{figure}[htb]
\begin{center}
\includegraphics[width=.45\linewidth]{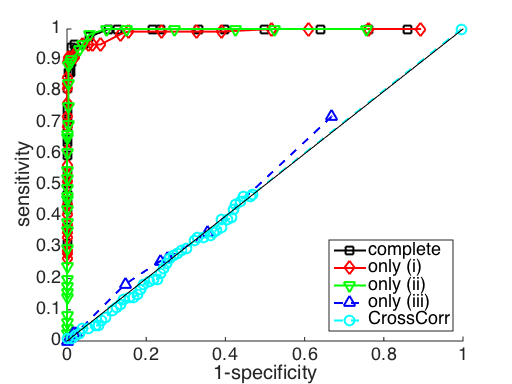}
\includegraphics[width=.45\linewidth]{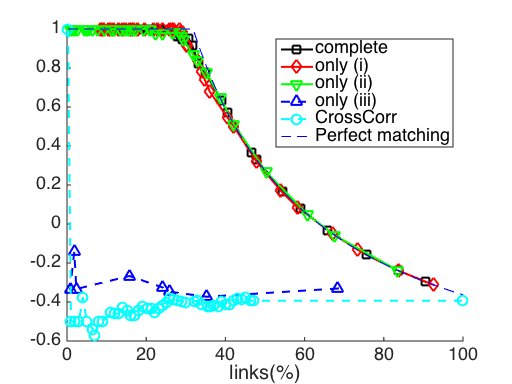}
\caption{ROC (on the left) and PPC (on the right) curves for the topology estimator \eqref{eq:Vhat}, with the complete data preprocessing and
with each of the filtering conditions (i), (ii) and (iii) given in Section~\ref{sec:preproc}.
The first conditions (i) and (ii) of data preprocessing may be used alone without affecting the main result,
while spiking activity alone is not predictive}\label{fig:ROC}
\end{center}
\end{figure}

The validation of the proposed methodology is afforded on simulations presented in the Section~\ref{simul}. The performances of the methodologies are quantified with the well known receiver operating characteristic (ROC) curves and with the recently introduced positive precision curve (PPC, \cite{Garofalo2009}). The graphs reported in Figure \ref{fig:ROC} are obtained by varying the penalization parameter $\lambda$, for the Lasso methodology, and the cross-correlation threshold, for the respective correlation methodology. 

Let us remind that PPC is defined as:
\begin{equation}
	\mathrm{PPC}=\frac{TP-FP}{TP+FP}
\end{equation}
and represents the proportion of the correctly (true positive, TP) versus the incorrectly (false positive, FP) inferred links. The plot on the right in the Figure \ref{fig:ROC} reports the PPC index with respect to the number of links included in the analysis. Interestingly, the PPC curve of the Lasso stays at its maximum up to 30\% of the links, that corresponds to the real links of the network (the connectivity probability is 0.3). 
Clearly, beyond 30\% of the included links, the PPC has a negative power decay. 
We point out that the cross-correlation does not achieve to infer the topology at any level of the cross-correlation threshold. Interestingly, when the Lasso events are based only on spiking information (condition (iii) alone) it performs as bad as the cross-correlation. 
The Lasso (Figure \ref{fig:ROC}) performs very well and reaches the maximum value (=1) in both the ROC and PPC curves for a large range of the penalization parameter. Moreover, the worst performances are achieved when no constraint is imposed on the sparsity (e.g. $\lambda=0$) of the inferred network. Finally, the intuition is preserved: the first connections introduced in the Lasso estimate, are 
almost all true ($\mathrm{PPC}\sim 1$).
Another interesting feature of the newly introduced Lasso methodology consists into its capability of inferring the birectional links. The results of Figure \ref{fig:ROC} are based on a network with 121 connections out of which 38 were bidirectional (31.4\% out of the total). This is an another advantage over the cross-correlation method, that only determines unidirectional connections.

\section{Conclusions and Future Works}
Understanding how interactions between brain structures support the performance of specific
cognitive tasks or perceptual processes is a prominent goal in neuroscience. The effects that one part of
the nervous system has on another are typically examined by stimulating or lesioning the
first part and investigating the outcome in the second. For example, in peripheral
and spinal pathways, the interventional techniques of stimulation and
ablation have proven to be powerful methods for inferring causal
influences from one neuron or neuronal population to another.
For the study of causal relations within the brain (functional
and effective connectivity ), the utility of the interventional techniques
is diminished by the high levels of convergence and divergence in brain pathways.\\
In this work we have developed an event based approach for inferring networks of causal relationships in a neuronal population.
Specifically, we suppose that we are able to observe the dynamical behaviors of individual components of a neuronal networks
and that few of the components may be causally influencing each other.  The variables could be invasive electrode recordings, intracranial EEG, or
non-invasive EEG, MEG or fMRI time series from different parts of the brain. In order to introduce our method we have considered a 
simulated cerebellar granule cell network capturing nonlinear interactions between even short and noisy time series. 

The results we got are quite promising from many point of views. 
First, despite the algorithm was applied only on short simulations (5 seconds) it achieved to filter out noisy from causal responses yielding a  reliable estimate of the underlying  connections in  the network. Second, the approach is quite general and the conditions (i),(ii),(iii) can be further adapted to different types of electro-physiological signals. Third, the Lasso is also quite robust respect to bidirectional connections in the network. This is of fundamental importance since bidirectional network motifs are quite abundant in the brain \cite{Song2005}.
The proposed Lasso methodology assumes the knowledge of the voltage traces. From an experimental point of view such a detailed information can be achieved with patch-clamp experiments but this technique is not designed to perform simultaneous recordings from populations of neurons.  
Interestingly, nowadays there have been huge improvements in the field of the genetically encoded voltage indicators that allow to track sub-threshold activities \cite{Gong2015}, a key ingredient of the Lasso algorithm. These progresses will likely provide in a near future multi-site recordings of a population of neurons.

As a next step, we will then test the robustness of the methodology including different noise sources (i.e. membrane noise), inhibitory connections and verify its robustness respect to bigger networks. 


\end{document}